\documentclass[12pt,a4paper]{iopart}

\usepackage{graphicx}
\usepackage{graphics}
\usepackage{setspace}
\usepackage{geometry}
\usepackage{xcolor}
\usepackage{subfig}
\usepackage{tikz}
\usepackage{placeins}
\usepackage{adjustbox}
\usepackage{multirow}
\usepackage[labelfont=bf]{caption}

\usepackage{soul}
\usepackage{color}

\begin{document}

\title[Inception networks, DA, and TL in photosensitivity diagnosis]{Inception networks, Data Augmentation and Transfer Learning in EEG-based photosensitivity diagnosis}

\author{Fernando Moncada Martins$^{1,3,*}$ and V{\'i}ctor M. Gonz{\'a}lez$^{1,3}$ and Jos{\'e} R. Villar$^{2,3}$ and Beatriz Garc{\'i}a L{\'o}pez$^4$ and Ana Isabel G{\'o}mez-Men{\'e}ndez$^4$}

\address{1 Electrical Engineering Department, University of Oviedo, Gij{\'o}n, Asturias, Spain. 33203}
\address{2 Computer Science Department, University of Oviedo. Gij{\'o}n, Asturias, Spain. 33203}
\address{3 Biomedical Engineering Center, University of Oviedo. Gij{\'o}n, Asturias, Spain. 33203}
\address{4 Clinical Neurophysiology Department, Burgos University Hospital. Burgos, Castilla y León, Spain. 09006}

\address{* Corresponding Author}

\eads{\mailto{moncadafernando@uniovi.es}$^{*}$, \mailto{vmsuarez@uniovi.es}, \mailto{villarjose@uniovi.es}, \mailto{bgarcialo@saludcastillayleon.es}, \mailto{agomm@saludcastillayleon.es}}

\begin{abstract}
Photosensitivity refers to a neurophysiological condition in which the brain generates epileptic discharges known as Photoparoxysmal Responses (PPR) in response to light flashes. In severe cases, these PPR can lead to epileptic seizures. The standardized diagnostic procedure for this condition is called Intermittent Photic Stimulation. During this procedure, the patient is exposed to a flashing light, aiming to trigger these epileptic reactions while preventing their full development. Meanwhile, brain activity is monitored using Electroencephalography, which is visually analyzed by clinical staff to identify these responses. Hence, the automatic detection of PPR becomes a highly unbalanced problem that has been barely studied in the literature due to photosensitivity’s low prevalence. 
This research tackles this problem and proposes using Inception-based Deep Learning (DL) neural networks that, together with transfer learning, are trained in epilepsy seizure detection and tuned in the PPR automatic detection task. A Data Augmentation (DA) technique is also applied to balance the available data set, evaluating its effects on the DL models. The proposal outperformed state-of-the-art solutions in the literature, achieving higher ratios on standard performance metrics, and with DA significantly improving the Sensitivity without affecting Accuracy and Specificity. This project is currently being developed with patients from Burgos University Hospital, Spain.
\end{abstract}

%
\vspace{2pc}

%
\maketitle
%
%


\section{Introduction}
\label{sec:int}
While epilepsy diagnosis is receiving most of the attention from the research community \cite{Hossain2019,Ibrahim2022,Chakrabarti2021}, photosensitivity is not that much under its focus.
Photosensitivity is a neurological condition defined as an abnormal responsiveness of the brain to certain visual stimuli, such as light reflections and patterns, able to trigger a paroxistic response in the form of epileptic discharges called Photoparoxysmal Responses (PPR) \cite{Dorothee1989,MeritamLarsen2021}. As stated in \cite{Dorothee1989}, up to 6\% of the healthy population suffer from photosensitivity, a percentage that raises to 30\% for epileptic patients \cite{Fisher2022}. Besides, there are four types of these epileptic phenomena defined in \cite{Waltz1992}  --from type I to IV; the higher the number, the greater the severity, and the higher the probability of provoking generalized seizures--. 
However,  it is hard to classify every real PPR into just one of these types because of the intrinsic mixture of the four types and the inherent differences among patients and conditions.

The clinical diagnosis of photosensitivity follows a standardized procedure known as Intermittent Photic Stimulation (IPS) \cite{Dorothee1999,Rubboli2004}, inciting the patient with intermittent flashes at different increasing --and then decreasing-- frequencies, while the Electroencephalogram (EEG) signals are continuously and on-line monitored. The stimulation ends whenever a PPR is detected, registering the frequency range of photosensitivity for the patient. In addition to the inherent IPS' well-known drawbacks \cite{Fisher2022,Strigaro2021,Parra2007} (such as the number of human resources needed or the lack of colour analysis), the procedure fails in producing a balanced data set. Effectively, the small percentage of affected people \cite{Dorothee1989,MeritamLarsen2021,Rathore2020}, together with the fact that the process ends when a PPR is detected, provoke that the gathered data includes an exceedingly short number of PPR signals, hindering the creation of a suitable data set for training Machine Learning (ML) and Deep Learning (DL) models \cite{Moncada2022Journal,Moncada2023Journal}.  

This research proposes using a Time Series-designed Inception-based model \cite{Ismail2020} to detect PPR activity in our photosensitivity data set. Transfer Learning (TL) tackles the lack of PPR data to satisfactorily train DL models by pretraining the Inception model using a  publicly available data set \cite{Physionet2000} recorded from clinical epileptic patients. A subsequent tuning stage adapts the obtained model to the PPR classification. In addition, an ad-hoc Data Augmentation (DA) technique enriches the photosensitivity data set with synthetic PPR Time Series instances. This research is part of a larger project introduced in \cite{Moncada2022Journal}, proposing the integration of Artificial Intelligence (AI) and Virtual Reality technologies into epilepsy --and more specifically, photosensitivity-- diagnosing process.

The structure of this paper is as follows: the next subsection introduces the related work. Section \ref{sec:mat_met} deals with the data set information, the method description, the experimental setup, and the evaluation scheme. Then, section \ref{sec:res} shows the results obtained from the experimentation along with the discussion about the outcomes and knowledge extracted. The paper ends by drawing the main conclusions from this research. 

\subsection{Related work}
AI techniques have been widely used for the automatic detection, classification, and prediction of epileptic seizures, from simple ML methods \cite{Siddiqui2020} to more complex DL models \cite{Shoeibi2021}. The catalog of recently employed AI techniques for EEG-based epilepsy seizure detection varies from Extreme Gradient Boosting \cite{Vanabelle2020}, K-Nearest Neighbors (KNN) and Fuzzy Rough Nearest Neighbors \cite{Zia2021}, and DL --with almost all the possible structures, such as dense layers \cite{Choubey2021},  Convolutional Neural Network (CNN) \cite{Tang2021},  Long Short-Term Memory (LSTM) \cite{Chakrabarti2021}, or autoencoders  \cite{Abdelhameed2021}--. 

Notwithstanding the research in this field, there is still room for improvement \cite{Rasheed2021_ML}. The study in \cite{Sahu2020} compared several of these ML techniques and CNN, finding that the latter outperformed the other epileptic seizure detection methods. Therefore, DL seems the path to follow for EEG pattern analysis; however, the need for a high volume of training and testing data becomes a challenge. Due to the restricted and limited nature of clinical and medical data, and the inherent imbalance of the data due to the IPS diagnostic procedure, the lack of suitable data sets becomes a serious challenge that needs addressing. 
Tackling the data availability and quality problems implies either DA, or TL, or both \cite{Ismail2018}.

On one hand, applying DA techniques to create synthetic data to increase the number of minority-class instances, i.e., computing very simple operations to data --Jittering, Time or Spectral Warping, Slicing or Scaling-- or using more complex generative models like Autoencoders or Generative networks, can produce realistic data from the original data set \cite{Iwana2021}. DA has already been tested in clinical problems such as fall detection with an LSTM-based model in \cite{EnolNeurocomputing2022}, improving electro-cardiogram (ECG) classification in \cite{Jacaruso2021}, single-channel EEG sleep stage classification in \cite{He2022,Khalili2021} or epileptic seizure prediction in \cite{Rasheed2021_DA,Shu2023}. 

On the other hand, TL enables the use of larger data sets from different, yet similar domains to that of the current problem to pre-train the AI models, learning the basic patterns and operations from the domain, and transferring them to the problem at hand by tuning the model with the target data set \cite{Pan2010}. As said, TL is one of those techniques that recently started to be a field of interest in time-series processing tasks \cite{Ismail2018}, including the medical field \cite{Gupta2019}, where it has already been tested for transferring features learned from image processing to ECG time-series classification \cite{Gajedran2021}, or to perform different EEG analysis \cite{Wan2021}, e.g., specific-patient motor imagery decoding using adaptive TL \cite{Zhang2021}, the prediction of clustered seizures in epileptic patients \cite{Cao2023}, or the detection and classification of seven seizure types \cite{Raghu2020}.

Nevertheless, to our knowledge, the automatic PPR detection problem has not yet been thoroughly addressed. The study in \cite{Strigaro2021} used the high-frequency brain oscillations (30--140Hz) evoked as a response to a stimulation procedure different from the clinical IPS: it consisted of white flashes at 0.3Hz, making it possible to analyze the brain activity from 50ms pre-stimulus to 400ms post-stimulus. This research found evidence that some features of these early reactions could be used as photosensitivity biomarkers. In contrast, the authors of \cite{Parra2002} proposed an analysis of Discrete Fourier Components extracted from EEG segments just before a PPR to test whether or not the spectral activity becomes handy in predicting these phenomena. This study defined the relative phase clustering index metric to compare frequency components and evaluate if any of them was stronger than the fundamental component (the one corresponding to the stimulation frequency of the IPS process). They found that this value greatly increases before the epileptic discharges, with potential use as an IPS predictor. A previous work \cite{Moncada2023Conf} modified this procedure to perform PPR detection rather than prediction, as it is what the neurophysiologists need to diagnose. 
This new approach divided each instance into smaller fragments to perform the spectral analysis instead of analysing the Fourier Components from a sequence of various EEG segments. However, the results did not preserve the original ones, meaning the new experimentation was unsuitable for the detection problem.

Multi-stage ML methods for automatic PPR detection were proposed in \cite{Moncada2022Journal,Moncada2023Journal,Moncada2022Conf}. 
Firstly, a one-class KNN classifier is used as an anomaly detector, evaluating the normality of an EEG window, and detecting any instance with abnormal brain activity. Secondly, a binary KNN analyzes all the abnormal EEG windows to infer if the anomaly is caused by PPR activity or not \cite{Moncada2022Journal}.  Additionally, the study in \cite{Moncada2022Conf} proposed a K-Means algorithm for clustering EEG windows, evaluating several classic ML classifiers in each cluster according to the ratio of PPR windows within them. Moreover, an ad-hoc DA technique that generates realistic PPR windows is proposed in \cite{Moncada2023Journal}. This method selects two PPR windows by splitting them into segments. Afterwards, a new PPR window is generated by alternatively choosing segments from each of the original windows and joining them one after the other; interpolating the points in the junctions to avoid the discontinuities in the signal. This DA approach improved the data set balance and the classification results of our previous best ML models.


\section {Material and Methods}
\label{sec:mat_met}

This section details the materials, the proposed methods, and the experimentation design. Firstly, Section \ref{subsec:data} describes the source and target data sets employed in the experimentation in this research and the preprocessing steps applied to both of them. Secondly, Section \ref{subsec:inception_transfer} introduces the architecture of the InceptionTime model, describing the ensemble of the trained networks. Afterwards, Section \ref{sec:TL} describes the Transfer Learning methodology. Following, Section \ref{sec:DA} deals with the details of the DA stage. Finally, Section \ref{subsec:exp} details the metrics and the experimental design.

\subsection{Data Sets and Pre-processing}
\label{subsec:data}

Transfer Learning requires two data sets: the source one --gathered from a close domain, definitively big enough for the DL training--, and the target one used for tuning in and evaluating the final model. 

\begin{figure}[htb] 
    \centering
    \subfloat{\includegraphics[width = 0.47\textwidth]{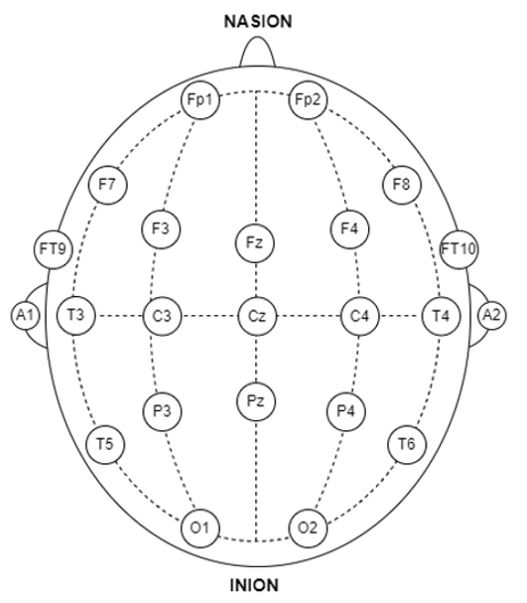}}
    \hspace{5mm}
    \subfloat{\includegraphics[width = 0.47\textwidth]{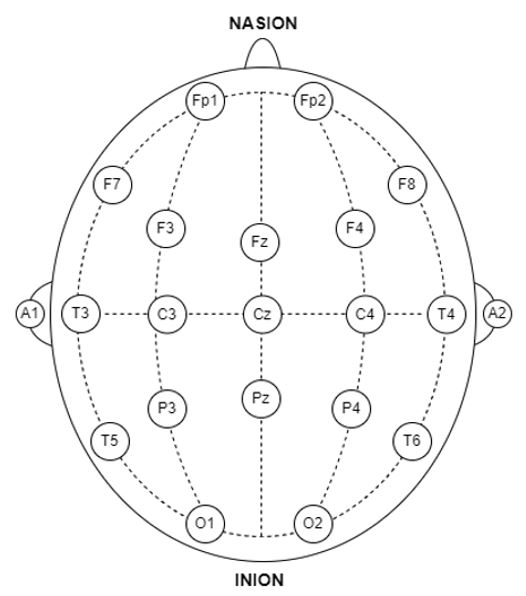}}
    \caption{Position of the electrodes following the 10-20 international standardized system: \textit{Epi\_Data} was recorded using 21 electrodes (\textbf{left)}, while \textit{Phot\_Data} used 19 electrodes (\textbf{right}). The \textit{Nasion} is located at the center of the frontonasal area, and the \textit{Inion} is at the center of the occipital area. A1 and A2 are the ground electrodes.} 
    \label{fig:eeg_channels}
\end{figure}

The source data set is the free and public CHB-MIT Scalp EEG Database (\textit{Epi\_Data} from now on) from Physionet \cite{Physionet2000}, reported in \cite{CHB-MIT2010}: it includes a collection of EEG recordings from $N_S=22$ patients that suffered epileptic seizures. Each patient was monitored for several days, resulting in 9-to-42 continuous EEG recordings of 1-to-4 hours in length per patient; this research only considers those sessions that include seizure recordings. Each EEG was recorded using a non-invasive EEG cap with 21 electrodes placed according to the 10-20 standardized system \cite{EEGCommittee1958} (as shown in the left part of Figure \ref{fig:eeg_channels}) at a sampling rate of 256Hz. A total of 182 seizures are annotated marking their starting and ending timestamps. They applied the bipolar montage, where the voltage of each electrode is subtracted from the voltage value of the channel behind it, forming a chain from front to back, resulting in a total of 18 combinations, plus 4 ad-hoc combinations --please, refer to the source paper for the available pairs--.

The clinical neurophysiologists of Burgos University Hospital recorded the target data set from photosensitive patients (\textit{Phot\_Data} from now on). It includes recordings from 10 photosensitivity-diagnosed patients. Each patient was submitted to a 3-to-5-minute continuous IPS session while recording their brain activity by EEG, each EEG was recorded using the Natus Nicolet v44 cap, which included 19 electrodes placed according to the 10-20 standardized system \cite{EEGCommittee1958} (as shown in the right part of Figure \ref{fig:eeg_channels}) at a sampling rate of 500Hz. The clinical specialists supervised each recording using the Natus Neuroworks $\copyright$ software to mark each PPR region triggered during the stimulation process using the average montage --the daily used montage--, which calculates the average voltage of all channels at each timestamp and subtracts it from the raw measured voltage of each channel.

The multivariate EEG signals were segmented using a 1-second length sliding window. However, each data set requires a different window overlap due to the excessive length of the recordings and the resulting unbalanced ratio. 
On the one hand, for the \textit{Phot\_Data}, with shorter recordings --about 5 minutes length-- and a reduced number of short-time PPR, this study proposes a window overlap of 90\%, so several windows may partly include the same PPR  --enhancing the richness of these PPR windows--. On the other hand, the  \textit{Epi\_Data} includes extremely long recordings --from 1 to 4 hours long--, with most of the time the patient exhibiting normal behaviour and only a few moments registering an epileptic seizure. In this case, a window overlap would only increment the unbalanced ratio; therefore,  this study opted not to apply any window overlap to this data set. The different overlapping values allow to avoid I) the possibility of cutting PPR discharges in separate windows and create instances that contain the whole PPR activity as well as its triggering or its closure in \textit{Phot\_Data}; and II) generating an excessive number of EEG instances that could accentuate the imbalance effect even more in \textit{Epi\_Data}.

As a summary, Table \ref{tab:data_prop} shows the obtained number of windows for each data set, with the unbalanced ratio. Also, this table shows the balanced ratio to reach on the Phot\_Data after a DA stage.

\begin{table}[hbt]
    \centering
    \begin{tabular}{c|c|c|cc}
         & \textit{Epi\_Data} & \textit{Phot\_Data} & \multicolumn{2}{c}{\textit{Phot\_Data + DA}}\\
         &      &  &  Train  & Test\\
         \hline
        Total number of windows & 671,299 & 29,190 & 7,500 & 2,900  \\
        Epileptic windows &  10,860 & 1,222 & 3,000 & 120 \\
        \% of anomaly windows &  1.62\% & 4.19\%  & 40.00\%  & 4.14\% \\
    \end{tabular}
    \caption{Distribution of epileptic windows in the \textit{Epi\_Data} and \textit{Phot\_Data} data sets, including the figures when using Data Augmentation. As explained in Section \ref{sec:EXP2}, DA generates synthetic windows until reaching the required number of PPR windows while reducing the number of non-PPR windows to 7.500 resampling windows from all the subjects included in the training fold.}
    \label{tab:data_prop}
\end{table}

\begin{figure}[htbp]
    \centering
    \includegraphics[width = 0.8 \textwidth]{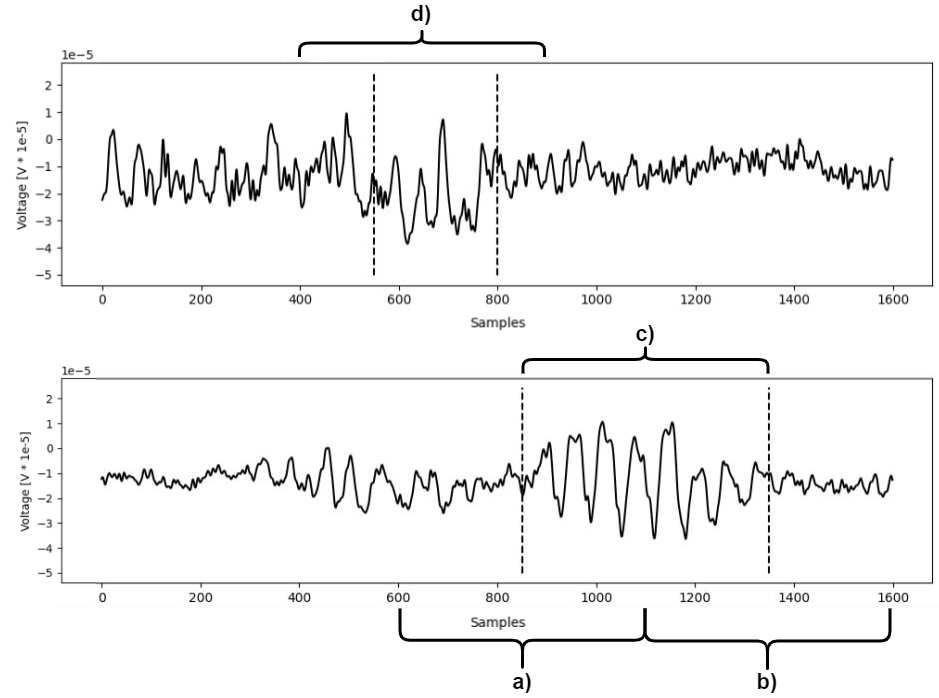}
    \caption{The sliding windowing generates four types of windows that include PPR signal segments. These figures depict two EEG channels that include one PPR each --delimited with dashed lines--. On the top one, when the PPR is a short event, the sliding window may enclose the whole phenomenon --type \textbf{(d)} window--. The bottom one shows the three remaining cases: windows that \textbf{(a)} only include the starting part of the PPR (onset); \textbf{(b)} only contains the final part of the PPR (offset); \textbf{(c)} are fully occupied by PPR signal (no onset nor offset).}
    \label{fig:typesPPRwind}
\end{figure}

Besides, using sliding windowing on the EEG signals from \textit{Phot\_Data} created four different ways the sliding window cuts PPR windows (refer to Figure \ref{fig:typesPPRwind}): \textbf{(a)} windows that include only a PPR starting point, \textbf{(b)} windows that present only a PPR ending point, \textbf{(c)} windows that coincide with the PPR activity, and \textbf{(d)} windows that contain a whole PPR due to their small duration (less than 1 second). Each PPR window is assigned to one of these four groups as required for the DA stage.

The differences in the protocols for recording the EEG sessions force extra pre-processing; this pre-processing focuses on unifying the channels, the montage, and the sampling rate. Considering the number of channels, both data sets used the 10-20 system, but \textit{Epi\_Data} has more channels (FT9 and FT10); consequently, the pre-processing omits these two channels. Moreover,  bipolar montage is the selected montage for both data sets, transforming the data representation in \textit{Phot\_Data} accordingly. Finally, \textit{Epi\_Data} recorded their EEG signals at 256Hz, while \textit{Phot\_Data} used 500Hz. {Cubic spline interpolation \cite{csi}} increases the number of samples in \textit{Epi\_Data}. 
    
Therefore, there are two data sets --the source and the target, with subscripts $S$ and $T$. On the one hand, \textit{Epi\_Data} includes data from $N_S=22$ patients, with a different number of recordings per patient, and each recording varying in length. A sliding window of 1-second length produces the source data set $\{X^S_i,y^S_i\}_{i=1}{^{m_S}}$, with $X^S_i \in R^{C_S\times T_S}$, $C_S=18$ channels, $T_S=500$ data points per window, $m_S=671,299$ total number of windows. Besides, $y^S_i \in \{SEIZURE,\ NORMAL\}$. On the other hand, \textit{Phot\_Data} includes data from $N_S=10$ patients, with a single recording per patient, and each recording varying in length. A sliding window of 1-second length produces the source data set $\{X^T_i,y^T_i\}_{i=1}{^{m_T}}$, with $X^T_i \in R^{C_T\times T_T}$, $C_T=18$ channels, $T_T=500$ data points per window, $m_T=29,190$ total number of windows. Besides, $y^T_i \in \{PPR,\ NORMAL\}$.

\subsection{Inception-based Model}
\label{subsec:inception_transfer}

\begin{figure}[hbt]
    \centering
    \includegraphics[width = 0.99\textwidth]{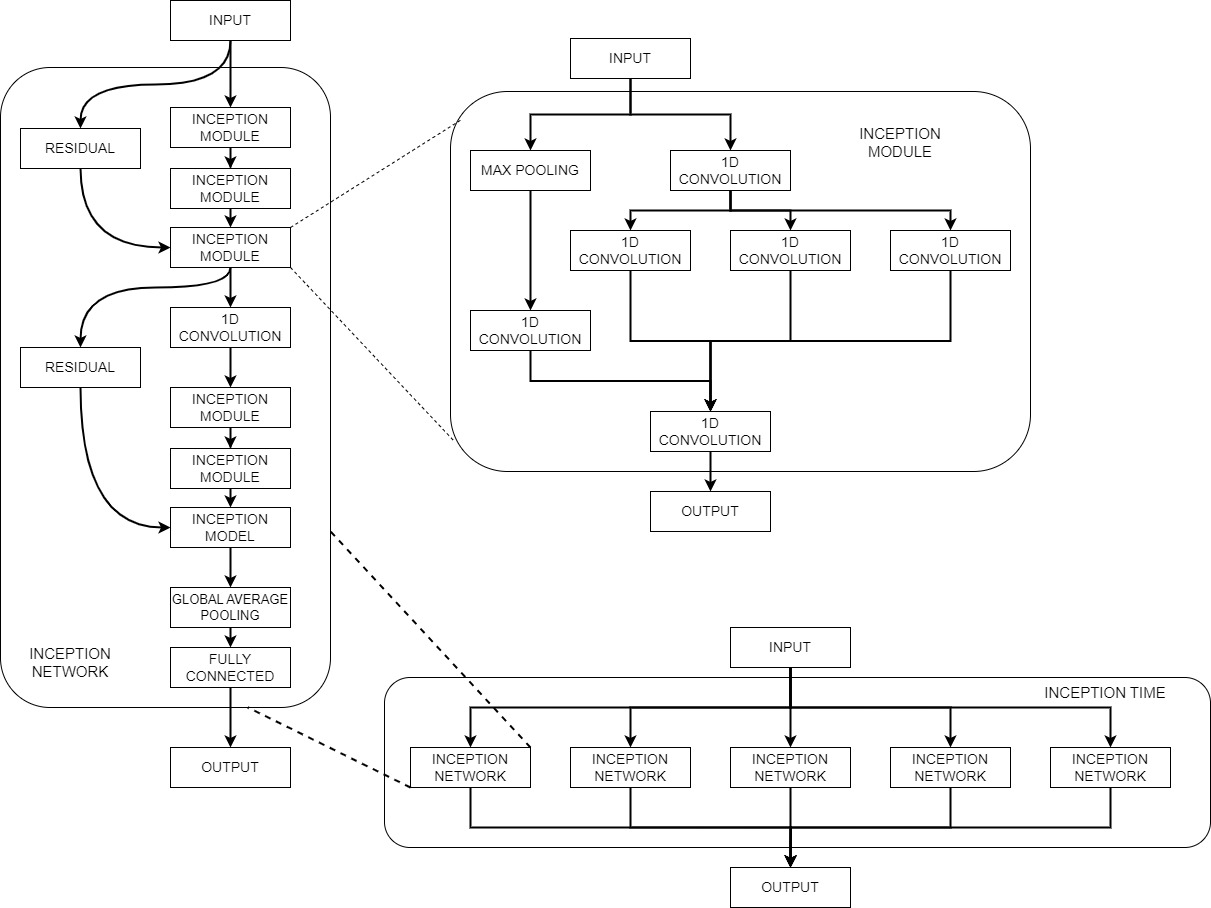}
    \caption{The architecture of one of the InceptionTime model, including 5 Inception Networks in parallel. The Inception Networks includes 3 Inception modules, which represent a simplification of the original Inception module \cite{inception}.}
    \label{fig:incep_time}
\end{figure}

This research proposes the InceptionTime DL neural network model for automatic PPR detection \cite{Ismail2020}, a Time Series classification variation of the Inception model. An Inception Network consists of a sequence of two residual blocks, followed by a global average pooling, and a fully connected soft-max layer. Each residual block includes a sequence of three Inception modules, with a residual layer that diverts the residual block's input to the last Inception network (see Figure  \ref{fig:incep_time}). 

Each Inception module includes a 1D-Convolutional layer acting as a bottleneck, whose output is connected to three side-by-side 1D-Convolutional layers of different sizes. In parallel, a Max-pooling layer receives the input signal, and its output is the input of a 1D-Convolutional layer of size 1. The outputs from all the convolutional layers are concatenated, acting as the input of a final 1D-Convolutional layer, which computes the output of the Inception module.

Finally, the InceptionTime model is the ensemble of 5 Inception Networks \cite{Ismail2020}; the authors explained this decision based on the high standard deviation in accuracy that relies on the weights initialization and the training process. Instead of going deeper on the network or training longer, the authors opted to create an ensemble of high-standard deviation Inception Networks. In this research, we want to study this variation as well, hence, we propose, on the one hand, to use a single Inception Network, and on the other hand, to use the ensemble of 5 Inception Networks conforming to the InceptionTime model.

The ensemble of classifiers is a simple average of the probabilistic values that each Inception Network calculates for each label, as stated in Eq. \ref{eq:ensemble}. In this equation, $\lambda$ represents one of the possible labels to assign, with $\lambda \in \Lambda$, and $\Lambda$ is the set of labels of the problem --\{SEIZURE, NORMAL\} for the \textit{Epi\_Data}, \{PPR, NORMAL\} for the \textit{Phot\_Data}--. Moreover, $p_{i,\lambda}$ is the probability of instance $X_i^D$ to belong to label $\lambda$, $D$ stands for $S$ or $T$ --for the source or target data set, respectively--. Finally, $\theta_j$ represents each Inception Network --with $j=1, \dots, n=5$--, while  $\theta_j\lambda$ is the logistic output of the corresponding Înception network $\theta_j$ for label $\lambda$.

\begin{equation}
    p_{i,\lambda}=\frac{1}{n}\times \sum_{j=1}^{n}{\theta_j^\lambda(X_i^D)}, \quad  \forall \lambda \in \Lambda \quad \land \quad n=5
    \label{eq:ensemble}
\end{equation}

\subsection{Transfer Learning Strategy}
\label{sec:TL}
\begin{figure}[!htb] 
    \centering
    \includegraphics[width =\textwidth]{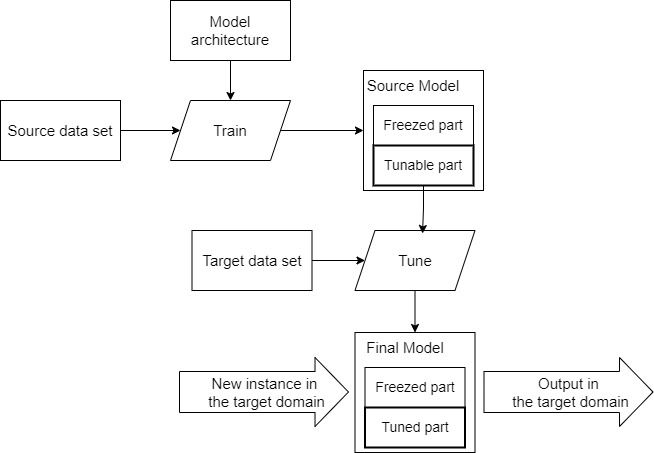}
    \caption{Concept of Transfer Learning: first, a model is trained with the source data to train the model and, more specifically, to learn the patterns and their processing (the model's freezer part). Then, when transferring the knowledge to the target domain, only the model's final layers (tunable part) are tuned with the target data for the target task.}
    \label{fig:transfer}
\end{figure}

TL proposes learning a model in one domain where data availability is not compromised, tuning some of its stages --usually, the final ones-- afterward with the scarce data from the target domain where the model will be exploited \cite{Pan2010,TransferLearning1976}; Figure  \ref{fig:transfer} illustrates this concept. The training and tuning data sets are denoted as \textbf{source} and \textbf{target} data, respectively. Typically, this latter data set is smaller than the former.

Among the different approaches to TL (inductive TL, transductive TL, or unsupervised TL), we can consider this research the former one because the source and target domains are similar, while the task at hand differs from one to the other: for the \textit{Epi\_Data}, the task is to classify a Time Series window as epileptic onset or not; for the \textit{Phot\_Data}, the task is to label a Time Series window as being a PPR or not.

The InceptionTime model will be trained as their authors explain in \cite{Ismail2020} using the \textit{Epi\_Data} as the source data set. The needed adaptations in the target data set (\textit{Phot\_Data}) have already been detailed in the previous sections, including increasing the sampling rate for the \textit{Epi\_Data}, or changing to a bipolar montage in the target data set. To this end, the trained Inception Networks are translated from the source to the target domain, tuning only the last two Inception modules, the Global Average Pooling layer and the last fully-connected layer --which is set to one neuron for binary classification--.

\subsection{Data Augmentation for balancing a PPR data set}
\label{sec:DA}

\begin{figure}[htb] 
    \centering
    \includegraphics[width = 0.85\textwidth]{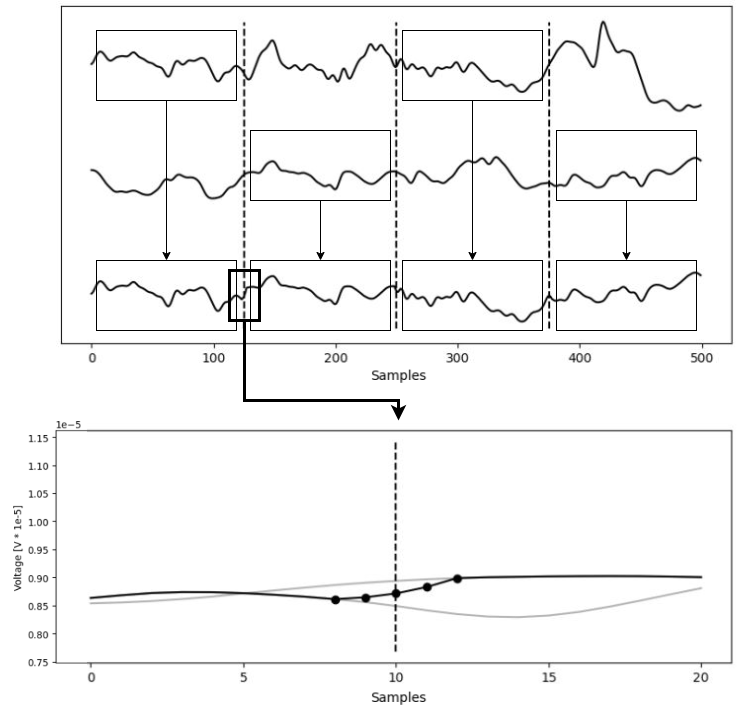}
    \caption{Example of the DA technique. The top part shows three signals: the two original channels and the synthetic one. The first and second rows are the two windows to merge. Vertical dashed lines mark the cut-points. Alternating segments from these two windows are selected to create the new synthetic window (third row). The bottom part illustrates the adaptation between two consecutive segments in the synthetic signal.}
    \label{fig:DA}
\end{figure}

In \cite{Moncada2023Journal}, a DA stage balanced the EEG data set by injecting realistic Time Series for the problem of PPR detection. Figure \ref{fig:DA} illustrates the DA process: it starts by choosing two random PPR windows --A and B-- of the same type and segmenting them into equal-length sections. Secondly, it builds up a new synthetic window by joining alternating segments from A and B. Finally, a last step ensures smooth transitions between two signals around the cut-points using a weighted sum of the V points around the cut-point (V/2 points before and V/2 points after it, V is an even integer); this stage helps filter the abrupt changes in the generated signal that might cause variations in the spectrogram of the synthetic Time Series.

The weights for the ending segment vary from 1.0 for the V/2 points before the cut-point to 0.0 for the V/2 points after it; the other way around happens with weight for the starting segment: weights go from 0.0 for V/2 points before the cut-point to 1.0 for V/2 points after it. Eq. \ref{eq:smoothing} defines the weighted sum of the two original segments ($A$ and $B$), with $x$ being the cut-point, where up to 5 points are adapted to smooth the transitions from $A$ to $B$. The bottom part of Figure \ref{fig:DA} visualises this segment merging. When more than one EEG channel is considered, the same cut-points are used among all the channels to keep coherency.

\begin{equation}
\begin{array}{rcl}
C(x-2) &=& \hspace{12mm} A(x-2) \\
C(x-1) &=& 0.75 * A(x-1) + 0.25 * B(x-1) \\ 
C(x)   &=& \hspace{0.2cm} 0.5 * A(x) \hspace{7mm} + 0.50 * B(x) \\ 
C(x+1) &=& 0.25 * A(x+1) + 0.75 * B(x+1) \\
C(x+2) &=& \hspace{44mm} B(x+2)  
\end{array}
\label{eq:smoothing}
\end{equation}

\subsection{Experimentation Design}
\label{subsec:exp}
The experimentation setup includes three differentiated parts. The first stage (EXP1, Section \ref{sec:EXP1}) studies the performance of Inception Networks, their variability, and the ensemble outcome. 
Afterwards, the second stage (EXP2, Section \ref{sec:EXP2}) measures how DA can benefit the tuning of the Inception Networks and the final ensemble. The last stage (EXP3, Section \ref{sec:EXP3}) compares the approach proposed in this study to state-of-the-art methods for PPR detection. The subsequent subsections deal with each of these aspects.

\subsubsection{EXP1: Performance of InceptionTime for PPR detection} \hfill \\
\label{sec:EXP1}

EXP1 aims to evaluate the performance of each of the 5 Inception Networks, plus the performance of the ensemble of these models. TL trains each Inception Network in the source domain using the \textit{Epi\_Data}, splitting the dataset in 90\% train and 10\% test. Then, a Leave-One-Subject-Out (LOSO) cross-validation splits the \textit{Phot\_Data} in training and testing, adjusting the tunable part of the Inception Network. Replicating this process for each of the 5 Inception Networks generates the complete InceptionTime, aggregating their outcomes.

\begin{figure}[htb] 
    \centering
    \includegraphics[width = 0.99\textwidth]{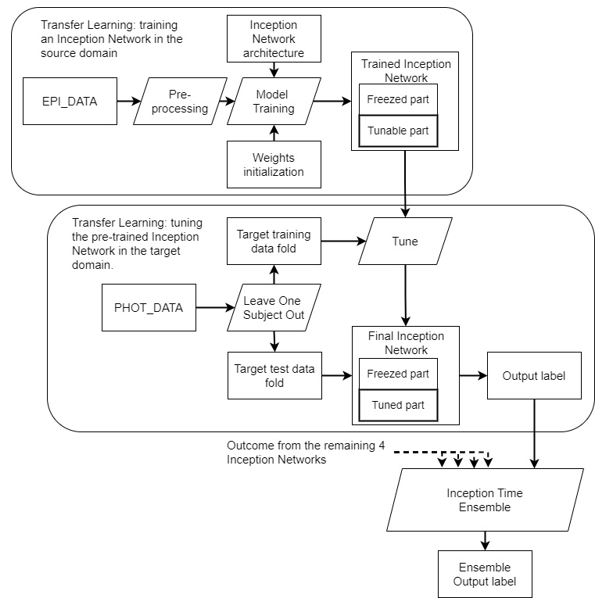}
    \caption{EXP1: Evaluation of the Inception Networks and InceptionTime model for PPR detection. The five Inception Networks are evaluated independently, but their outcomes are also aggregated as an ensemble following \cite{Ismail2020}.}
    \label{fig:exp1_workflow}
\end{figure}

Figure  \ref{fig:exp1_workflow} depicts the workflow of this experimentation: training in the source domain, transferring the model and tuning it in the target domain, and calculating the ensemble of the five replications of the Inception Networks. This Figure only shows one Inception Network training and tuning for clarity.

This experimentation produces the outcome of each Inception Network for every patient in the \textit{Phot\_Data}. Thus, the aggregation of these networks with the ensemble is also available for every patient. Hence, it is possible to compare the performance of the Inception Networks and the InceptionTime.

\subsubsection{EXP2: Effects of DA in InceptionTime model} \hfill \\
\label{sec:EXP2}

EXP2 evaluates the relevance of DA in tuning the Inception Networks in the \textit{Phto\_Data}data set. It replicates the same experimentation as in the previous section but introduces the DA stage in the tuning part of the TL. The data from each patient includes a total amount of 2900 EEG windows, from which around 120 are PPR instances, on average (as stated in Table \ref{tab:data_prop}). 
When following the LOSO scheme, each train set is created with around 26100 EEG windows, only 900 being PPR activity. To improve the balance of the training set, a DA step is applied to increase the number of PPR instances up to 3000; then, a random undersampling reduces the number of normal instances until reaching a 60\%/40\% proportion of non-PPR/PPR instances, i.e., 4500 normal instances in a set of 7500. On the other hand, the test set maintains all the original data.

\begin{figure}[htb] 
    \centering
    \includegraphics[width = 0.99\textwidth]{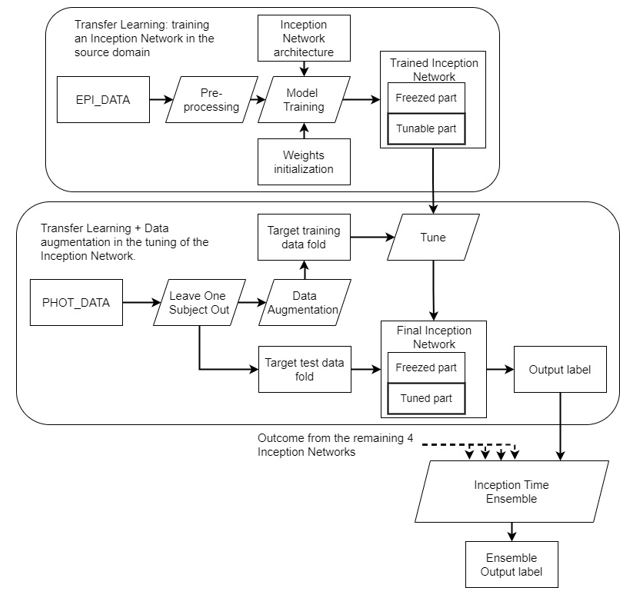}
    \caption{EXP2 scheme: Evaluation of the effects of DA in the models' performance.}
    \label{fig:exp2_final_workflow}
\end{figure}

The remaining parts of the experimentation remain the same, with no changes at all. Figure \ref{fig:exp2_final_workflow} shows the whole process, with the only modification due to the inclusion of the DA stage. 
As before, this experimentation not only compares the individual performance of each Inception Network but also visualizes the improvements in the InceptionTime ensemble due to the inclusion of the DA stage.

\subsubsection{EXP3: A comparison of PPR detection methods} \hfill \\
\label{sec:EXP3}

Finally, this experimental setup compares the proposed approach with the best state-of-the-art methods for PPR detection; to our knowledge, the study in \cite{Moncada2023Journal,Moncada2022Conf} represents the best approach for PPR detection published so far, where a Dense-Layer Neural Network (DL-NN) defeated Support Vector Machines and K-Nearest Neighbor --either in the form of one-class or binary classifiers--, as well as Random Forests for PPR detection. A pre-processing stage extracted several features from each window --such as statistical features (Kurtosis, Skewness, etc.), temporal domain features (Sum of Absolute Values, etc.), and spectral domain features (Maximum Power
Spectrum, Spectral Centroid, Spectral Density, etc.)--; afterwards, Principal Component Analysis reduced the dimension of the problem to 12 input features.

\begin{figure}[htb] 
    \centering
    \includegraphics[width = \textwidth]{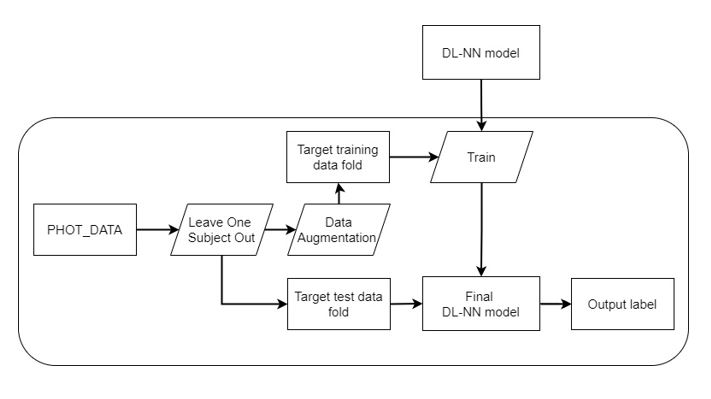}
    \caption{EXP3: Training the DL-NN from \cite{Moncada2022Journal,Moncada2022Conf} using the same experimental setup proposed in this research.}
    \label{fig:exp3_final_workflow}
\end{figure}

Consequently, this experimental setup applies the same LOSO scheme to train the DL-NN classifier --refer to Figure  \ref{fig:exp3_final_workflow}--, so it becomes possible to compare its performance results to those of the Inception Networks and InceptionTime model. For this purpose, the number of neurons of the hidden layer of the DL-NN is determined, evaluating different possibilities --10, 20, 30, 40, and 50 neurons--.

\subsection{Evaluation and results visualization}
\label{subsec:eval}

Accuracy (ACC, Eq. \ref{eq:acc}), Sensitivity (SENS, Eq. \ref{eq:sens}) and Specificity (SPEC, Eq. \ref{eq:spec}) are the metrics that measure the performance of the methods in the three experimental setups; in these equations, TP, TN, FP and FN stand for the True Positive, True Negative, False Positive and False Negative counters, respectively.  Due to the inherent unbalanced problem, these three metrics will provide a good insight into what is happening with the different methods.

\begin{equation}
\label{eq:acc}
    ACC = \frac{TP + TN}{TP + TN + FP + FN}
\end{equation}

\begin{equation}
\label{eq:sens}
    SENS = \frac{TP}{TP + FN}
\end{equation}

\begin{equation}
\label{eq:spec}
    SPEC = \frac{TN}{TN + FP}
\end{equation}

To avoid overloading the reader with tables and figures from each experimental setup, only one table and a single boxplot will condense all the results. The table will include the mean, median and standard deviation among all the patients in the LOSO cross-validation. Furthermore, a boxplot will represent the dispersion in the test values for every model. The table and boxplot will include results from the Inception Network and InceptionTime with and without DA and from the DL-NN model. Nevertheless, a final Annex will contain the complete tabulated results for each patient.


\section {Results and Discussion}
\label{sec:res}

\begin{figure}
    \centering
    \includegraphics[width = \textwidth,trim={0mm 24 0 19},clip]{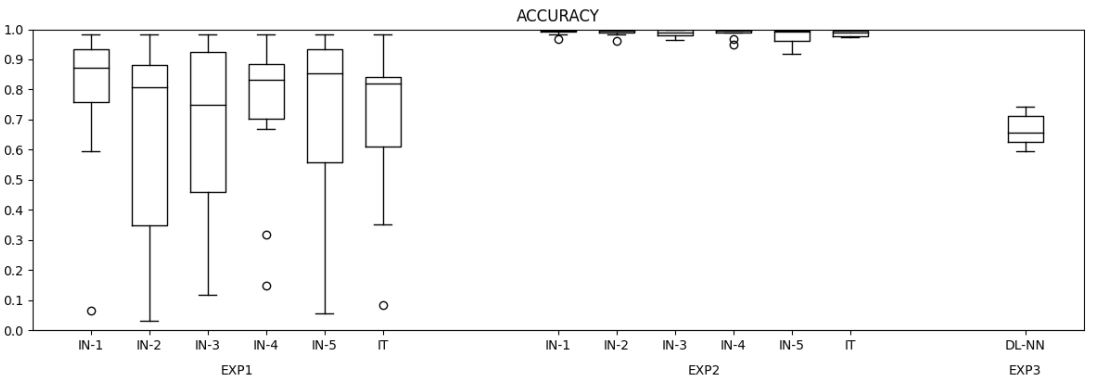} \\ 
    \includegraphics[width = \textwidth,trim={0mm 20 0 19},clip]{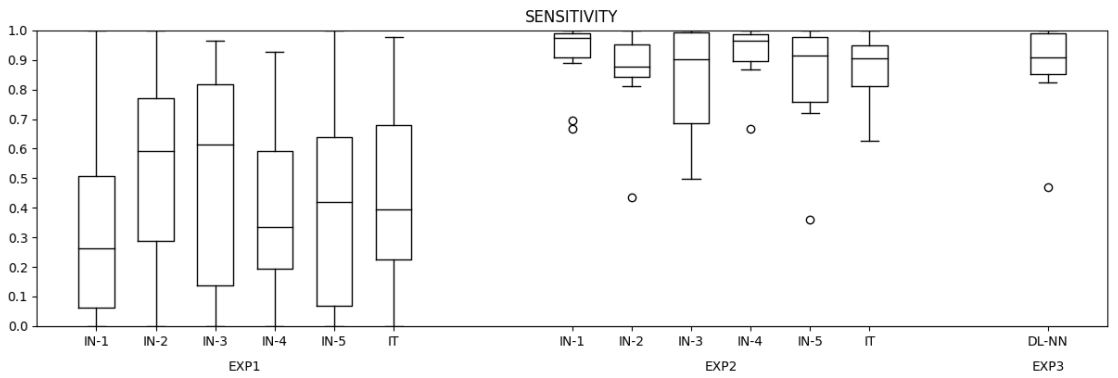} \\
    \includegraphics[width = \textwidth,trim={0mm 0 0 19},clip]{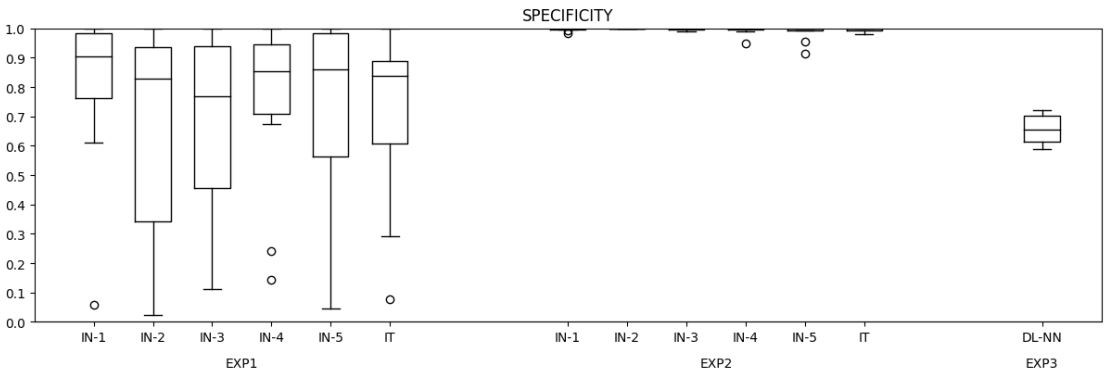} \\
    \caption{Boxplots of the results arranged by metric: ACC, SENS, and SPEC are depicted in the top, middle, and bottom parts. The boxplots show, from left to right, i) the 5 Inception Networks plus the InceptionTime model without DA (EXP1), ii) the 5 Inception Networks plus the InceptionTime model with DA (EXP2), and iii) the best DL-NN model with DA (EXP3).}
    \label{fig:EXP2_DA_results}
\end{figure}

Table \ref{table:brief_results} summarizes the mean (Mn), median (Md) and standard deviation (StD) of the performance metrics for all the experiments: the 5 Inception Networks plus the InceptionTime model following EXP1, the 5 Inception Networks plus the InceptionTime model using DA in the training (EXP2), and the best DL-NN model obtained from EXP3. Furthermore, Figure \ref{fig:EXP2_DA_results} depicts each metric's boxplot for all the experiments. \ref{appendix1} includes the whole set of results, tabulating them in different tables --one per experiment--, containing the results for each subject. Boxplots of the results from EXP3 for the different evaluated DL-NN models are also presented.

\begin{table}
    \centering
    \resizebox{\textwidth}{!}{
        \begin{tabular}{c|c c c c c | c || c c c c c | c || c}
        & \multicolumn{6}{c ||}{InceptionTime} & \multicolumn{6}{c||}{InceptionTime + DA} & \\
        & IN-1 & IN-2 & IN-3 & IN-4 & IN-5 & IT & IN-1 & IN-2 & IN-3 & IN-4 & IN-5 & IT & DL-NN\\
        \hline
        \multicolumn{14}{c}{\textbf{ACC}} \\ 
        \hline
                                    
        Mn   & 0.7764 & 0.6328 & 0.6675 & 0.7295 & 0.7158 & 0.7044 & \textbf{0.9918} & 0.9902 & 0.9868 & 0.9875 & 0.9776 & \textbf{\textit{0.9868}} & 0.6653 \\ 
        Md & 0.8722 & 0.8077 & 0.7491 & 0.8320 & 0.8532 & 0.8205 & \textbf{{0.9965}} & 0.9940 & 0.9908 & 0.9954 & 0.9911 & \textbf{\textit{0.9884}} & 0.6569 \\
        StD    & 0.2619 & 0.3231 & 0.2940 & 0.2630 & 0.3007 & 0.2739  & \textbf{{0.0097}} & 0.0110 & 0.0126 & 0.0161 & 0.0253 & \textbf{\textit{0.0103}} & 0.0531 \\  
    
        \hline
       
        \multicolumn{14}{c}{\textbf{SENS}} \\ 
        \hline
   
        Mn   & 0.3523 & 0.5139 & 0.4975 & 0.3905 & 0.4074 & 0.4324 & 0.9149 & 0.8580 & 0.8247 & \textbf{0.9236} & 0.8439 & 0.8730 & \textbf{\textit{0.8792}} \\
        Md & 0.2620 & 0.5935 & 0.6143 & 0.3356 & 0.4186 & 0.3938 & \textbf{0.9736} & 0.8787 & 0.9007 & 0.9662 & 0.9155 & 0.9046 & \textbf{\textit{0.9084}} \\
        StD    & 0.3250 & 0.3292 & 0.3605 & 0.3008 & 0.3269 & 0.3092 & 0.1206 & 0.1543 & 0.1934 & \textbf{0.0963} & 0.1886 & \textbf{\textit{0.1106}} & 0.1503 \\
    
        \hline
    
        \multicolumn{14}{c}{\textbf{SPEC}} \\ 
        \hline
          
        Mn   & 0.7993 & 0.6388 & 0.6680 & 0.7420 & 0.7286 & 0.7154 & 0.9967 & \textbf{0.9993} & 0.9978 & 0.9927 & 0.9856 & \textbf{\textit{0.9944}} & 0.6561 \\ 
        Md & 0.9037 & 0.8271 & 0.7676 & 0.8549 & 0.8609 & 0.8385 & 0.9995 & \textbf{0.9998} & 0.9993 & 0.9993 & 0.9993 & \textbf{\textit{0.9981}} & 0.6564 \\
        StD    & 0.2748 & 0.3524 & 0.3204 & 0.2902 & 0.3219 & 0.2957 & 0.0053 & \textbf{0.0008} & 0.0029 & 0.0153 & 0.0275 & \textbf{\textit{0.0069}} & 0.0484 \\ 
    
        \hline
        \end{tabular}
         } 
    \caption{Summary of the results. Mean (Mn), Median (Md), and Standard Deviation (StD) of the ACC (top part), SENS (middle part), and SPEC (bottom part) metrics. Columns, from left to right, correspond to a) the 5 Inception Networks (IN) and the InceptionTime (IT) model from EXP1, b) the 5 IN and the IT model with DA from EXP2, and c) the best DL-NN model found in EXP3. Bold letters remark the best value so far; in bold and italics, the best final model --among the InceptionTime models and the DL-NN model--.} 
    
    \label{table:brief_results}
    
\end{table}

Perhaps the most remarkable result at a first glance to Table \ref{table:brief_results} is the poor results obtained for the Inception Networks and the InceptionTime model with no DA. Not only low-performance metrics are observed, but also a rather high variability in the performance from one subject to another. Moreover, the obtained SENS values for several subjects are even worse than flipping a coin. There were even patients (P1 and P2) whose epileptic phenomena remained completely undetected as represented by a SENS value of 0\%. The concern about the variability in the results was already mentioned in \cite{Ismail2020} without affecting the InceptionTime model; however, with the challenging data set in EXP1, the InceptionTime model failed to perform similarly for all the subjects.

The performance of the DL-NN model after being trained with DA is also striking. The DL-NN model detects PPR patterns more effectively than any other window type, suggesting some level of overfitting. However, this effect was not found in the training, so there must be no reason for this performance. Nevertheless, the SPEC reduces so much with respect to the results in \cite{Moncada2023Journal,Moncada2022Conf}, making this solution unfeasible. This variability in the SPEC and ACC is relatively worse because the percentage of PPR windows for each subject is so small: the number of false alarms is too high. The performance deterioration is mainly due to the difference in the experimentation, using a 10-fold-like cross-validation used in the cited studies, while LOSO --used in this research, with a closer resemblance to what might happen when deploying the models-- produces a more realistic vision of the performance of the models.  By the way, these results could have been expected for DL-NN but the other way around --higher SPEC and lower SENS--. 

Finally, the effect of applying DA in the training of the Inception Networks and InceptionTime models becomes noticeable: from a wrong and invalid variability in the metrics to an almost unnoticeable variability of the SPEC and ACC. As explained before, the high percentage of non-PPR windows forces that high SPEC induces high ACC, and small variabilities in the SPEC induce slight variations in the ACC. Hence, although IN3 trained with DA has a relatively high variability in the SENS, the variability in the ACC for this Inception Network remains almost constant. DA creates such an amount of synthetic PPR instances that further surpasses the quantity of real PPR instances. This fact could result in the appearance of a drift in the data, affecting the final results. If more real patients were evaluated using the proposed methodology, misclassification is expected to be stronger, worsening the detection results.
In addition, it is possible to appreciate the effects of the InceptionTime model in merging the SENS results from each Inception Network and producing a final ensemble with slight variations in performance; this behaviour was claimed by the authors in \cite{Ismail2020}, but is only noticeable when DA is applied to balance the training subsets.

The favourable effects of the DA suggest this strategy needs further development, such as using Generative models that would lead to higher generalization capabilities. Additionally, different architectures, such as Transformers networks with Attention mechanisms, would be evaluated and compared to define the best possible solution. 
Besides, these obtained results may suggest a change in the paradigm of the problem. Perhaps, unsupervised learning and anomaly detection represent a better approach instead of a supervised learning process. A recent study \cite{moncadaCAEPIA2024} proposes this research line, which may become interesting due to the inherent imbalance of the problem.


\FloatBarrier

\section {Conclusions}
\label{sec:conc}

In this study, we proposed using an InceptionTime model for PPR activity detection in EEG recordings from a photosensitivity data set; an InceptionTime model is the ensemble of Inception Networks. TL applies from training the models with a well-known Epilepsy EEG seizure detection data set to adjusting the models on the photosensitivity domain. A  suitable preprocessing stage guarantees the compatibility of both data sets, splitting the EEG recordings into 1-second length windows. Furthermore, a DA stage balances the training data set in the target domain; the DA produces realistic synthetic PPR windows by merging two similar PPR windows and introducing weighting mechanisms that avoid abrupt changes.

The experimentation considers LOSO cross-validation, comparing the InceptionTime models trained with and without DA versus a DL-NN model previously reported in the literature \cite{Moncada2023Journal,Moncada2022Conf}. Results show that using DA enhances the InceptionTime model, obtaining almost constant performances in terms of SPEC and ACC while keeping high SENS values. The comparison also showed that InceptionTime defeats the DL-NN for PPR detection. However, due to the high imbalance existing in \textit{Phot\_Data}, the high performance could seem a bit illogical when following a LOSO strategy, even after applying DA techniques to improve the balance of the training sets. The reason may be the creation of a domain shift produced by the large number of synthetic windows created during the balancing phase. In order to grasp the main issues underneath this glaring phenomenon, more in-depth study and analysis are required.

Results also suggest different future works. For instance, Generative models should allow for reaching higher generalization capabilities. Moreover, architectures --such as Transformers networks with Attention mechanisms-- would lead to better performance models. Finally, changing the problem's paradigm to unsupervised and anomaly detection may help to enhance the results.



\ack{This research has been funded by the Spanish Ministry of Economics and Industry, grant PID2020-112726RB-I00, the Spanish Research Agency --grant PID2023-146257OB-I00--, and by the Ministry of Science and Innovation under Missions Science and Innovation project MIG-20211008 (INMERBOT). Also, by Principado de Asturias, grant SV-PA-21-AYUD/2021/50994, and by the Council of Gijón through the University Institute of Industrial Technology of Asturias {grants SV-23-GIJON-1-09 and SV‐24‐GIJON‐1‐05. Finally, this research has also been funded by Fundaci\'{o}n Universidad de Oviedo grants FUO-23-008 and FUO-22-450.}} \\

\noindent {\small \textbf{Conflicts of interest} \\
The authors declare that they have no known competing financial interests or personal relationships that could have appeared to influence the work reported in this paper. } \\

\noindent {\small \textbf{Ethical approval} \\
The study was conducted in accordance with the Declaration of Helsinki and approved by the Ethics Committee of the Burgos University Hospital (Protocol Code CEIm2467, 23 February 2021). \\
Informed consent was obtained from all subjects involved in the study. Written informed consent was obtained from the patients to publish this paper.} \\

\noindent {\small \textbf{Author Contributions} \\
Fernando Moncada Martins, Víctor M. González, and José R. Villar designed the methodology and experiments and executed all technical and computer work. They all wrote the paper. \\
Clinical specialists Beatriz García López and Ana Isabel Gómez-Menéndez collected the data at Burgos University Hospital. They looked for suitable EEG recordings from photosensitive patients with an appropriate amount of PPR, { anonymized and labled them.} \\

\noindent {\small \textbf{Abbreviations} \\
The following abbreviations are used in this manuscript: \\ \\
\begin{tabular}{@{}ll}
ACC & Accuracy \\
AI & Artificial Intelligence \\
CNN & Convolutional Neural Network \\
DA & Data Augmentation \\
DL & Deep Learning \\
DL-NN & Dense-Layer Neural Network \\
ECG & Electro-cardiogram \\
EEG & Electro-encephalogram \\
IPS & Intermittent Photic Stimulation \\
KNN & K-Nearest-Neighbors \\
LOSO & Leave-One-Subject-Out \\
LSTM & Long-Short Term Memory \\
Md & Median \\
ML & Machine Learning \\
Mn & Mean \\
PPR & Photoparoxysmal Response \\
SENS & Sensitivity \\
SPEC & Specificity \\
StD & Standard Deviation \\
TL & Transfer Learning \\
\end{tabular}
}

\section*{References}

\bibliographystyle{unsrt}
\bibliography{bibliography}

\newpage

\appendix
\section{Result Tables}
\label{appendix1}

\begin{table}[!htb] 
    \centering
    \resizebox{0.55\textwidth}{!}{
        \begin{tabular}{c|c c c c c | c}
        $P_i$ & IN-1 & IN-2 & IN-3 & IN-4 & IN-5 & IT \\
        \hline
        \hline
        \multicolumn{7}{c}{\textbf{ACC}} \\ 
        \hline

        $P_1$ & 0.9522 & 0.9529 & 0.9515 & 0.9386 & 0.9525 & 0.9496 \\
        $P_2$ & 0.9831 & 0.9831 & 0.9829 & 0.9831 & 0.9831 & 0.9831 \\
        $P_3$ & 0.8838 & 0.8900 & 0.6669 & 0.8823 & 0.8823 & 0.8411 \\
        $P_4$ & 0.8320 & 0.8529 & 0.6935 & 0.8302 & 0.8780 & 0.8173 \\
        $P_5$ & 0.9739 & 0.3245 & 0.9448 & 0.8842 & 0.9787 & 0.8212 \\
        $P_6$ & 0.0656 & 0.0322 & 0.1178 & 0.1494 & 0.0544 & 0.0839 \\
        $P_7$ & 0.8706 & 0.8111 & 0.8662 & 0.8057 & 0.8284 & 0.8364 \\
        $P_8$ & 0.7344 & 0.4193 & 0.3921 & 0.6702 & 0.4380 & 0.5398 \\
        $P_9$ & 0.8738 & 0.8043 & 0.8046 & 0.8338 & 0.7826 & 0.8198 \\
        $P_{10}$ & 0.5944 & 0.2564 & 0.2548 & 0.3180 & 0.3354 & 0.3518 \\
        \hline                                     
        Mean   & 0.7764 & 0.6328 & 0.6675 & 0.7295 & 0.7158 & 0.7044 \\ 
        Median & 0.8722 & 0.8077 & 0.7491 & 0.8320 & 0.8532 & 0.8205 \\
        StD    & 0.2619 & 0.3231 & 0.2940 & 0.2630 & 0.3007 & 0.2739 \\  
    
        \hline
        \hline

        \multicolumn{7}{c}{\textbf{SENS}} \\ 
        \hline
 
        $P_1$ & 0.0000 & 0.0000 & 0.0000 & 0.0000 & 0.0000 & 0.0000 \\
        $P_2$ & 0.0000 & 0.0000 & 0.0000 & 0.0000 & 0.0000 & 0.0000 \\
        $P_3$ & 0.2045 & 0.2652 & 0.7424 & 0.1894 & 0.2727 & 0.3348 \\
        $P_4$ & 0.7794 & 0.6176 & 0.8382 & 0.6176 & 0.5882 & 0.6882 \\
        $P_5$ & 0.0204 & 0.7143 & 0.1020 & 0.2041 & 0.0000 & 0.2082 \\
        $P_6$ & 1.0000 & 1.0000 & 0.9643 & 0.9286 & 1.0000 & 0.9786 \\
        $P_7$ & 0.3194 & 0.5694 & 0.4861 & 0.4028 & 0.4861 & 0.4528 \\
        $P_8$ & 0.5116 & 0.7907 & 0.7674 & 0.5116 & 0.7209 & 0.6605 \\
        $P_9$ & 0.1902 & 0.3512 & 0.2390 & 0.2683 & 0.3512 & 0.2800 \\
        $P_{10}$ & 0.4977 & 0.8310 & 0.8356 & 0.7824 & 0.6551 & 0.7204 \\
        \hline                                              
        Mean   & 0.3523 & 0.5139 & 0.4975 & 0.3905 & 0.4074 & 0.4324 \\
        Median & 0.2620 & 0.5935 & 0.6143 & 0.3356 & 0.4186 & 0.3938 \\
        StD    & 0.3250 & 0.3292 & 0.3605 & 0.3008 & 0.3269 & 0.3092 \\
    
        \hline
        \hline
    
        \multicolumn{7}{c}{\textbf{SPEC}} \\ 
        \hline

        $P_1$ & 0.9968 & 0.9975 & 0.9961 & 0.9826 & 0.9972 & 0.9940 \\
        $P_2$ & 1.0000 & 1.0000 & 0.9997 & 1.0000 & 1.0000 & 0.9999 \\
        $P_3$ & 0.9606 & 0.9606 & 0.6584 & 0.9606 & 0.9512 & 0.8983 \\
        $P_4$ & 0.8332 & 0.8586 & 0.6900 & 0.8353 & 0.8849 & 0.8204 \\
        $P_5$ & 0.9892 & 0.3183 & 0.9584 & 0.8951 & 0.9944 & 0.8311 \\
        $P_6$ & 0.0574 & 0.0236 & 0.1103 & 0.1425 & 0.0460 & 0.0760 \\
        $P_7$ & 0.8843 & 0.8172 & 0.8757 & 0.8158 & 0.8369 & 0.8460 \\
        $P_8$ & 0.7376 & 0.4140 & 0.3868 & 0.6724 & 0.4795 & 0.5381 \\
        $P_9$ & 0.9230 & 0.8369 & 0.8453 & 0.8745 & 0.8137 & 0.8587 \\
        $P_{10}$ & 0.6104 & 0.1616 & 0.1589 & 0.2414 & 0.2827 & 0.2910 \\
        \hline  
        Mean   & 0.7993 & 0.6388 & 0.6680 & 0.7420 & 0.7286 & 0.7154 \\ 
        Median & 0.9037 & 0.8271 & 0.7676 & 0.8549 & 0.8609 & 0.8385 \\
        StD    & 0.2748 & 0.3524 & 0.3204 & 0.2902 & 0.3219 & 0.2957 \\ 
    
        \hline
        \end{tabular}
        }
    \caption{EXP1 results: InceptionTime model applying LOSO without DA. ACC (top), SENS (middle), and SPEC (bottom) results for the 5 Inception Networks (IN) and the InceptionTime (IT) model.}
    
    \label{table:EXP1_LOSO_results}
    
\end{table}

\begin{table}[!htb] 
    \centering
    \resizebox{0.55\textwidth}{!}{
        \begin{tabular}{c|c c c c c | c}
        $P_i$ & IN-1 & IN-2 & IN-3 & IN-4 & IN-5 & IT \\
        \hline
        \hline
        \multicolumn{7}{c}{\textbf{ACC}} \\ 
        \hline

        $P_1$ & 0.9977 & 0.9838 & 1.0000 & 0.9977 & 0.9977 & 0.9954 \\
        $P_2$ & 0.9973 & 0.9970 & 0.9970 & 0.9973 & 0.9983 & 0.9974 \\
        $P_3$ & 0.9821 & 0.9952 & 0.9897 & 0.9478 & 0.9887 & 0.9807 \\
        $P_4$ & 0.9905 & 0.9607 & 0.9649 & 0.9908 & 0.9570 & 0.9728 \\
        $P_5$ & 0.9993 & 1.0000 & 1.0000 & 0.9959 & 0.9935 & 0.9977 \\
        $P_6$ & 0.9990 & 0.9941 & 0.9666 & 0.9948 & 0.9180 & 0.9745 \\
        $P_7$ & 0.9961 & 0.9938 & 0.9918 & 0.9961 & 0.9944 & 0.9944 \\
        $P_8$ & 0.9969 & 0.9984 & 0.9978 & 0.9997 & 0.9991 & 0.9984 \\
        $P_9$ & 0.9669 & 0.9859 & 0.9799 & 0.9662 & 0.9722 & 0.9742 \\
        $P_{10}$ & 0.9922 & 0.9935 & 0.9804 & 0.9891 & 0.9567 & 0.9824 \\
        \hline                                     
        Mean   & 0.9918 & 0.9902 & 0.9868 & 0.9875 & 0.9776 & 0.9868 \\ 
        Median & 0.9965 & 0.9940 & 0.9908 & 0.9954 & 0.9911 & 0.9884 \\
        StD    & 0.0097 & 0.0110 & 0.0126 & 0.0161 & 0.0253 & 0.0103 \\  
    
        \hline
        \hline

        \multicolumn{7}{c}{\textbf{SENS}} \\ 
        \hline

        $P_1$ & 0.9764 & 0.8346 & 1.0000 & 0.9764 & 0.9764 & 0.9528 \\
        $P_2$ & 0.8889 & 0.8750 & 0.8750 & 0.8889 & 0.9306 & 0.8917 \\
        $P_3$ & 0.9706 & 0.8824 & 0.9265 & 0.9559 & 0.8529 & 0.9176 \\
        $P_4$ & 0.9707 & 0.4341 & 0.4976 & 0.8683 & 0.3610 & 0.6263 \\
        $P_5$ & 1.0000 & 1.0000 & 1.0000 & 1.0000 & 0.7206 & 0.9441 \\
        $P_6$ & 0.9122 & 0.9122 & 0.5024 & 0.9902 & 0.9951 & 0.8780 \\
        $P_7$ & 0.9650 & 0.9650 & 0.9697 & 0.9767 & 0.9767 & 0.9753 \\
        $P_8$ & 1.0000 & 1.0000 & 1.0000 & 1.0000 & 1.0000 & 1.0000 \\
        $P_9$ & 0.6670 & 0.8667 & 0.8430 & 0.6676 & 0.7253 & 0.7539 \\
        $P_{10}$ & 0.6965 & 0.8103 & 0.6332 & 0.9118 & 0.9003 & 0.7904 \\
        \hline                                              
        Mean   & 0.9149 & 0.8580 & 0.8247 & 0.9236 & 0.8439 & 0.8730 \\
        Median & 0.9736 & 0.8787 & 0.9007 & 0.9662 & 0.9155 & 0.9046 \\
        StD    & 0.1206 & 0.1543 & 0.1934 & 0.0963 & 0.1886 & 0.1106 \\
    
        \hline
        \hline
    
        \multicolumn{7}{c}{\textbf{SPEC}} \\ 
        \hline

        $P_1$ & 1.0000 & 1.0000 & 1.0000 & 1.0000 & 1.0000 & 1.0000 \\
        $P_2$ & 1.0000 & 1.0000 & 1.0000 & 1.0000 & 1.0000 & 1.0000 \\
        $P_3$ & 0.9824 & 0.9979 & 0.9912 & 0.9476 & 0.9919 & 0.9822 \\
        $P_4$ & 0.9919 & 0.9986 & 0.9986 & 0.9996 & 1.0000 & 0.9978 \\
        $P_5$ & 0.9993 & 1.0000 & 1.0000 & 0.9958 & 1.0000 & 0.9990 \\
        $P_6$ & 0.9996 & 1.0000 & 1.0000 & 0.9951 & 0.9125 & 0.9814 \\
        $P_7$ & 0.9973 & 0.9985 & 0.9985 & 0.9992 & 0.9973 & 0.9976 \\
        $P_8$ & 0.9968 & 0.9984 & 0.9984 & 0.9997 & 0.9991 & 0.9984 \\
        $P_9$ & 0.9996 & 0.9996 & 0.9996 & 0.9994 & 0.9994 & 0.9986 \\
        $P_{10}$ & 0.9998 & 1.0000 & 1.0000 & 0.9911 & 0.9562 & 0.9894 \\
        \hline  
        Mean   & 0.9967 & 0.9993 & 0.9978 & 0.9927 & 0.9856 & 0.9944 \\ 
        Median & 0.9995 & 0.9998 & 0.9993 & 0.9993 & 0.9993 & 0.9981 \\
        StD    & 0.0053 & 0.0008 & 0.0029 & 0.0153 & 0.0275 & 0.0069 \\ 
    
        \hline
        \end{tabular}
    }
    \caption{EXP2 results: InceptionTime model applying LOSO and DA. ACC (top), SENS (middle), and SPEC (bottom) results for the 5 Inception Networks (IN) and the InceptionTime (IT) model.} 
    \label{table:EXP2_LOSO+DA_results}
\end{table}

\begin{table}[!htb] 
    \centering
    \resizebox{0.6\textwidth}{!}{
        \begin{tabular}{c|c c c c c}
        $P_i$ & DL-NN-10 & DL-NN-20 & DL-NN-30 & DL-NN-40 & DL-NN-50  \\
        \hline
        \hline
        \multicolumn{6}{c}{\textbf{ACC}} \\ 
        \hline

        $P_1$ & 0.6814 & 0.7139 & 0.7366 & 0.7231 & 0.6773 \\ 
        $P_2$ & 0.6287 & 0.6644 & 0.6365 & 0.6837 & 0.6920 \\
        $P_3$ & 0.7718 & 0.7378 & 0.7649 & 0.7417 & 0.7773 \\
        $P_4$ & 0.6536 & 0.6481 & 0.6787 & 0.6711 & 0.6619 \\
        $P_5$ & 0.5952 & 0.6281 & 0.6839 & 0.6426 & 0.6568 \\
        $P_6$ & 0.5978 & 0.6075 & 0.5953 & 0.5984 & 0.5875 \\
        $P_7$ & 0.6240 & 0.6115 & 0.6287 & 0.6253 & 0.6358 \\
        $P_8$ & 0.5784 & 0.5898 & 0.5777 & 0.5954 & 0.5820 \\
        $P_9$ & 0.6318 & 0.6252 & 0.6059 & 0.6292 & 0.6357 \\
        $P_{10}$ & 0.7649 & 0.7446 & 0.7745 & 0.7423 & 0.7580 \\
        \hline                                     
        Mean   & 0.6528 & 0.6571 & 0.6683 & 0.6653 & 0.6664 \\ 
        Median & 0.6303 & 0.6381 & 0.6576 & 0.6569 & 0.6593 \\
        StD    & 0.0643 & 0.0533 & 0.0675 & 0.0531 & 0.0607 \\  
    
        \hline
        \hline

        \multicolumn{6}{c}{\textbf{SENS}} \\ 
        \hline
        
        $P_1$ & 1.0000 & 1.0000 & 1.0000 & 1.0000 & 1.0000 \\
        $P_2$ & 1.0000 & 1.0000 & 1.0000 & 1.0000 & 0.9836 \\
        $P_3$ & 0.9280 & 0.8720 & 0.8560 & 0.9280 & 0.7760 \\
        $P_4$ & 0.9706 & 0.9853 & 0.9706 & 0.8235 & 0.5000 \\
        $P_5$ & 0.5102 & 0.4490 & 0.4286 & 0.4694 & 0.4694 \\
        $P_6$ & 0.9643 & 0.8214 & 1.0000 & 0.9643 & 0.6071 \\
        $P_7$ & 0.9167 & 0.9028 & 0.8889 & 0.8889 & 0.8889 \\
        $P_8$ & 1.0000 & 1.0000 & 0.6977 & 1.0000 & 0.2326 \\
        $P_9$ & 0.8634 & 0.8537 & 0.8780 & 0.8439 & 0.8634 \\
        $P_{10}$ & 0.9019 & 0.8925 & 0.8593 & 0.8738 & 0.8388 \\
        \hline                                              
        Mean   & 0.9055 & 0.8777 & 0.8580 & 0.8792 & 0.7160 \\
        Median & 0.9461 & 0.8977 & 0.8835 & 0.9084 & 0.8074 \\
        StD    & 0.1390 & 0.1569 & 0.1687 & 0.1503 & 0.2399 \\
    
        \hline
        \hline
    
        \multicolumn{6}{c}{\textbf{SPEC}} \\ 
        \hline
        
        $P_1$ & 0.6664 & 0.7005 & 0.7243 & 0.7101 & 0.6622 \\ 
        $P_2$ & 0.6224 & 0.6586 & 0.6302 & 0.6783 & 0.6870 \\
        $P_3$ & 0.7551 & 0.7235 & 0.7551 & 0.7217 & 0.7774 \\
        $P_4$ & 0.6460 & 0.6400 & 0.6717 & 0.6675 & 0.6657 \\
        $P_5$ & 0.5965 & 0.6309 & 0.6880 & 0.6454 & 0.6598 \\
        $P_6$ & 0.5946 & 0.6056 & 0.5917 & 0.5952 & 0.5873 \\
        $P_7$ & 0.6167 & 0.6042 & 0.6222 & 0.6188 & 0.6295 \\
        $P_8$ & 0.5723 & 0.5840 & 0.5760 & 0.5896 & 0.5870 \\
        $P_9$ & 0.6151 & 0.6088 & 0.5863 & 0.6137 & 0.6193 \\
        $P_{10}$ & 0.7426 & 0.7204 & 0.7605 & 0.7208 & 0.7448 \\
        \hline  
        Mean   & 0.6428 & 0.6477 & 0.6606 & 0.6561 & 0.6620 \\ 
        Median & 0.6195 & 0.6355 & 0.6510 & 0.6564 & 0.6610 \\
        StD    & 0.0587 & 0.0485 & 0.0661 & 0.0484 & 0.0591 \\ 
    
        \hline
        \end{tabular}
    }
    \caption{EXP3 results: DL-NN architecture applying LOSO and DA. ACC (top), SENS (middle), and SPEC (bottom) results for the 5 Networks with different number of neurons in the hidden layer.}
    \label{table:EXP3_LOSO_DLNN_results}
\end{table}

\begin{figure}
    \centering
    \includegraphics[width = \textwidth]{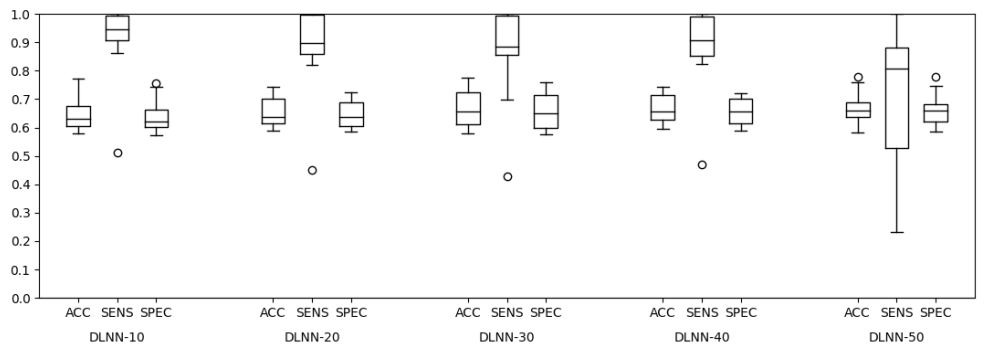}
    \caption{EXP3 results: boxplot of for each tested DL-NN model. TACC, SENS, and SPEC boxplots are shown for each case.}
    \label{fig:EXP3_LOSO_DLNN_boxplot}
\end{figure}

\end{document}